
\def\doublespace{\baselineskip=20pt plus 2pt\lineskip=3pt minus
     1pt\lineskiplimit=2pt}

\def\gsim{\mathrel{\raise.3ex\hbox{$>$}\mkern-14mu\lower0.6ex\hbox{$\sim$}}}
\def\lsim{\mathrel{\raise.3ex\hbox{$<$}\mkern-14mu\lower0.6ex\hbox{$\sim$}}}

\magnification=\magstep1
\hoffset 0.4truein
\hsize 6.2truein
\hfuzz 10pt
\doublespace

\centerline{\bf EJECTION OF FRAGMENTS IN SUPERNOVA EXPLOSIONS}
\bigskip
\centerline{\bf A.\ Loeb$^{\star}$, F.\ A.\ Rasio$^{\dag}$, \& J.\
Shaham$^{\ddag}$}
\bigskip
\noindent
$^{\star}$ Astronomy Department, Harvard University, Cambridge, MA 02138, USA
\smallskip
\noindent
$^{\dag}$ Institute For Advanced Study, Princeton, NJ 08540, USA
\smallskip
\noindent
$^{\ddag}$ Physics Department, Columbia University, New York, NY 10027, USA
\medskip
\hrule
\medskip

\noindent
{\bf Recent observations by the ROSAT X-ray satellite of the
Vela supernova remnant$^1$
have revealed, in addition to the previously identified compact nebula$^2$,
a nearly circular emitting region with a radius of about 4~degrees.
The Vela pulsar is slightly off the center of this circular region,
consistent with its measured proper motion$^{3}$
of about $100\,$km$\,$s$^{-1}$ and an age of about $10^4\,$yr.
The emitting region is bounded by the main supernova shock.
Just outside the shock, the X-ray image reveals several well-defined
V-shaped features extending radially outwards$^1$.
These features are most likely wakes produced
by objects moving supersonically through the outside medium.
The shapes and orientations of the wakes suggest that these objects
have been ejected from the center of the supernova explosion.
Their present positions indicate that they have been moving
with a mean velocity of a few thousand km$\,$s$^{-1}$.
We show that pre-existing objects, such as planets in orbit around the
progenitor star,
could not have been accelerated to sufficiently high velocities
or would have been destroyed. Instead, we propose that
the observed objects are fragments ejected during the formation
of the neutron star. Fragmentation during gravitational collapse
is a natural consequence of both convective$^4$ and rotational$^{5}$
instabilities.}

The X-ray emitting wakes observed by ROSAT
appear in the North, East and West of the Vela supernova remnant
and are stretched radially outwards$^1$.
Their angular distance from the center is about $4^{\circ}$,
giving a mean velocity
$v_f=2.8\times10^3 d_4\,\tau_4^{-1}\,{\rm km}\,{\rm s}^{-1}$,
where $400\,d_4\,$pc is the distance to the remnant and
$10^4\tau_4\,$yr is its age$^3$.
The initial ejection velocity must clearly have been $>v_f$.
For the brighter of the three features, the ROSAT PSPC countrate is
about 0.25$\,$s$^{-1}$ and the emitting area is about 32$\,$arcmin$^2$,
giving an X-ray luminosity $L_x\sim10^{34} d_4^2\,{\rm erg}\,{\rm s}^{-1}$
(J.\ Tr\"umper, personal communication). This implies a total radiated
energy of $\sim 3 \times 10^{45} d_4^2\,\tau_4~ {\rm erg}$ over the lifetime of
the remnant. If a moving fragment were to supply this energy while
decelerating from
the above $v_f$, its mass should be $\gsim 10^{-4} M_{\odot}$.

We can immediately rule out the possibility that the
fragments are pre-existing objects (e.g. planets)
that were accelerated by the explosion.
The progenitor star was most likely a supergiant of
radius $R_p>10^{13}\,$cm. The total energy imparted to
the ejected stellar envelope in a supernova explosion is $E_s\sim10^{51}\,$erg.
This energy includes both thermal and bulk velocity components.
The initial velocity of the shocked fluid  is $v_s\sim10^{4}\,{\rm km}\,{\rm
s}^{-1}$.
Thus, the total momentum flowing out is $<2E_s/v_s\sim 2\times 10^{42}\,
{\rm g}\,{\rm cm}\,{\rm s}^{-1}$.
The fraction of this momentum outflow that can be intercepted
by an object of radius $R_f$ outside the progenitor is $<(R_f/R_p)^2$.
Objects orbiting inside the progenitor are dragged to the center and
destroyed on a very short timescale$^{6}$ $\tau_d\sim 10^3\,$s.
The ejection velocity for an external object
with mass $M_f$ and mean density $\rho_f$ is
$v_f<[2E_s/(v_s M_f)](R_f/R_p)^2$, giving
$$v_f<6\, {\rm {km~ s^{-1}}}
\left({M_f\over 10^{-3}\,M_\odot}\right)^{-1/3}\left({\rho_f\over 1\,{\rm g\,
cm}^{-3}}\right)^{-2/3}\left({R_p\over 10^{13}\,{\rm cm}}\right)^{-2}  .
\eqno(1)
$$
For planets this upper limit is at least three orders of
magnitude smaller than the value of $v_f$ deduced from the observations.
Even if the planets could somehow survive inside the supergiant
envelope, the explosion would destroy them.
Indeed, disruption occurs whenever the nonuniform pressure forces exerted
by the ejected gas exceed a significant fraction
$\sim0.1$ of the internal gravitational force that keeps the planet bound.
A planet at a distance $r$ away from the center must be accelerated to
its ejection velocity $>v_f$ in a time $r/v_s$, and the condition
$v_f/(r/v_s)< 0.1{GM_f/R_f^2}$ translates to the lower bound
$$
r> 10^{15}\,{\rm cm} \left({M_f\over 10^{-3}\,M_\odot}\right)^{-1/3}
\left({\rho_f\over 1\,{\rm g}\,{\rm cm^{-3}}}\right)^{-2/3}
{\left({v_f\over 2\times 10^3 {\rm km}\,{\rm s^{-1}}}\right)}  .
\eqno(2)
$$
Expressions~(1) and~(2) show clearly that planets
cannot be accelerated to the required velocity without being totally
disrupted.

Instead we propose that the observed wakes are associated with
fragments that were ejected from inside the progenitor core during an
asymmetric gravitational collapse. During the acceleration phase, the
density inside the fragments was $\rho_i\sim10^{10-15}\,{\rm g}
\,{\rm cm}^{-3}$, high enough for the fragments to avoid disruption.
Direct evidence for asymmetric core collapse is provided by
the large observed proper motions of young pulsars$^7$.
Two specific mechanisms can result in the formation of small fragments.
Rotationally-induced instabilities during gravitational collapse$^5$ can lead
to
the development of either an axisymmetric, self-gravitating disk$^8$, or
an ellipsoidal deformation, which then leads to mass
shedding through outgoing spiral arms$^{9,10}$.
The spiral arms can later fragment
through a sausage instability, leaving small compact objects orbiting the
central
neutron star core$^{11}$.
The fragments in this case would be initially at close to nuclear density,
$\rho_i\sim10^{14}\,{\rm g}\,{\rm cm}^{-3}$,
and ejected from $r_i\sim10^{1-2}$km. They would be formed just outside the
radius
where the bounce shock forms initially, and so they could be shock-accelerated
and ejected before the shock stalls$^{12}$.
Alternatively, Rayleigh-Taylor (convective) instabilities in the outer iron
core$^4$
can lead to small blobs of overdense material forming
behind the shock front. These overdense blobs could be accelerated
by absorbing a small fraction of the neutrino flux, i.e., through the same
mechanism
that is now commonly thought to re-energize the stalled shock and power the
ejection
of the stellar envelope$^{4,12}$.
In this case the initial density is lower, $\rho_i\sim10^{10}\,{\rm g}\,{\rm
cm}^{-3}$,
and the ejection is from $r_i\sim 10^{2-3}\,$km.

In response to the release of nuclear energy in their interior,
fragments that are too small may disintegrate.
Consider a fragment made of hot ($T\gsim1\,$MeV), highly neutronized material.
As the fragment expands, nucleons recombine quickly into nuclei. At relatively
late times
the energy release is dominated by $\beta$-decays$^{13}$.
The maximum amount of nuclear energy released is about $8\,$MeV per nucleon if
all the matter transforms into iron.
The actual energy which gets thermalized inside the fragment is likely to be
considerably smaller,
because part of the nuclear energy is carried away by neutrinos,
and because the fragment may not form out of highly neutronized matter.
Ignoring degeneracy pressure, a necessary
condition for a fragment to remain gravitationally bound is that the
thermalized
nuclear energy be smaller than the gravitational binding energy
per baryon,
$$
T< E_{\rm grav}= 4\,{\rm MeV} \left({M_f\over 10^{-2}M_\odot}\right)^{2/3}
\left({\rho_i\over 10^{14}{\rm g~ cm^{-3}}}\right)^{1/3}  .
\eqno(8)
$$
We therefore obtain a conservative estimate of
$\sim0.01M_\odot$ for the minimum mass of a fragment that would not
disintegrate as a result of nuclear energy release.

Even if it can sustain the nuclear energy release,
a fragment could still be unbound if it is produced
with an initial temperature much larger than the virial
temperature. Just like a nascent neutron star,
a newly formed hot fragment can cool rapidly by neutrino
emission, to which it is optically thin. When the neutrino cooling
time becomes comparable to the hydrodynamic
time, the fragment begins expanding and may become unbound if its
temperature is still too high. The neutrino cooling time
due to pair annihilation is$^{14}$
$10^{-4}\,{\rm s} \left({T/{\rm 10\,MeV}}\right)^{-5}$.
If we take the sound speed to be $\sim 0.1c$
and the initial fragment size to be $\sim1\,$km, then the hydrodynamic
expansion time is $\sim 3\times10^{-5}\,$s. This becomes
comparable to the neutrino cooling time
at $T= T_c\approx 12\,$MeV. The exact value of the sound speed is not
important since $T_c\propto c_s^{1/5}$. Using equation (8)
we conclude that for initially hot fragments, the mass must be
$\gsim0.01 M_\odot$ to avoid disintegration.
This estimate turns out to be comparable to that obtained
above by considering only the nuclear energy release.

As the fragments move away from the center of the explosion on ballistic
orbits, they expand and eventually settle into
a gravitational equilibrium with a density $\sim 1\,{\rm g}\,{\rm cm^{-3}}$.
However, they could remain hot for a time $\gg10^4\,$yr
because of their large initial thermal heat content as well as
the energy release from radioactive nuclei in their interiors.
An evaporation process should result.
We write the evaporation rate in terms of the mass flux at the surface,
$$
{-\dot M_f}=4\pi \rho_e R_f^2 v_{e},
\eqno(10)
$$
where $\rho_e$ is the density of the wind at ejection and
where the ejection speed $v_e$ must be larger
than the escape velocity from the surface of the fragment,
$$
v_{e}
\gsim v_{esc}\equiv 1.3\times 10^{2}~{\rm km~ s^{-1}}
\left(M_f\over 10^{-2}M_\odot\right)^{1/3}
\left({\rho_f\over
1\,{\rm g}\,{\rm cm^{-3}}}\right)^{1/6}  .
\eqno(11)
$$
{}From mass conservation the density in the terminal velocity outflow is
$$
\rho(r)\approx {\rho_e R_f^2\over r^2} .
\eqno(12)
$$
We can estimate the radius $r_d$ where the outflow will be deflected
by the ram pressure of the external medium from the condition
$$
\rho (v_e^2-v_{esc}^2)= \rho_{ext} v_f^2  ,
\eqno(13)
$$
where $\rho_{ext}\sim 10^{-24}{\rm g~ cm^{-3}}$ is the
ambient interstellar density.
Taking $(v_{e}^2- v_{esc}^2)\sim v_{esc}^2$, we find
$$
r_d\approx\left({{\vert{\dot M_f}\vert}
v_{esc}\over 4\pi \rho_{ext} v_f^2}\right)^{1/2}
\sim 4\times10^{16}\,{\rm cm}\left({{\dot M_f}\over
{\dot M}_{max}}\right)^{1/2}
\left({v_f\over 2\times 10^3\,{\rm km}\,{\rm s^{-1}}}\right)^{-1}  ,
\eqno(14)
$$
where ${-{\dot M}_{max}}\equiv 10^{-6}M_\odot\,{\rm yr}^{-1}$ is the maximum
evaporation rate allowed for a fragment of mass $10^{-2}M_\odot$
over the current pulsar lifetime.
This radius $r_d$ gives the effective cross-section for estimating the
drag force $F_d\sim \rho_{ext} v_f^2 \times (\pi r_d^2)$. The
corresponding energy dissipation rate $F_d v_f$ yields a luminosity,
$$
L_x\approx \pi \rho_{ext} v_f^3 r_d^2 \sim 4\times 10^{34}\,{\rm erg}\,{\rm
s}^{-1}
\left({v_f\over 2\times 10^3\, {\rm km\,s^{-1}}}\right)
\left({{\dot M_f}\over {\dot M}_{max}}\right).
\eqno(15)$$
This crude estimate comes very close to the observed X-ray luminosity
determined from the ROSAT observation. Most of the energy is
expected to be released by thermal bremsstrahlung in X-rays since the
post-shock temperature of the gas is high. In fact, the Vela supernova may have
ejected
additional high velocity fragments. However, fragments with
$\vert{{\dot M}_f}\vert\ll
\vert{{\dot M}_{max}}\vert$
are too faint to be detectable by ROSAT, while those
with a much higher evaporation rate
would have disappeared by now.

To power an evaporation rate close to ${\dot M}_{max}$ requires
little energy deposition at the surface of a fragment,
namely $\sim 100\, {\rm eV}$ per baryon.
This energy can be easily supplied over $10^4$ years
by short-lived radioactivity or
by the initial heat content of the fragment. In order to maintain
an appreciable evaporation rate, the temperature at the surface of the
fragment must be $\sim 1/3$ of the escape temperature$^{15}$,
$$
T_{s}\approx 3\times 10^5{\rm K}\left({M_f\over 10^{-2}M_\odot}\right)^{2/3}
\left({\rho_f\over 1\,{\rm g}\,{\rm cm^{-3}}}\right)^{1/3}  .
\eqno(16)
$$
The material in the evaporative outflow is initially fully ionized.
As the surface material expands and rarefies to become part of the supersonic
wind,
it cools adiabatically with $T\propto \rho^{2/3}$. The gas starts to recombine
within a distance $\sim R_f$ and eventually reaches a sufficiently low
ionization
level, so that the outflow becomes
optically-thin, at a temperature $T_{ph}\approx 5000 {\rm K}$
around $r_{ph}\sim 3 R_f$.
The emerging optical luminosity from the wind photosphere is then,
$$
L_{opt}\approx 4\pi r_{ph}^2 \sigma T_{ph}^4 \sim
10^{33}\, {\rm erg}\,{\rm s^{-1}} \left({r_{ph}\over 3R_f}\right)^2
\left({M_f\over 10^{-2}M_\odot}\right)^{2/3}
\left({\rho_f\over 1\,{\rm g}\,{\rm cm^{-3}}}\right)^{-2/3}  \,
\eqno(17)
$$
where $\sigma$ is the Stefan-Boltzmann constant.
This estimate is highly uncertain because of the unknown
contribution of heavy elements and molecular absorption
to the opacity of the wind. Nevertheless, it shows
that the optical
luminosity of the fragments
may be detectable.
It should therefore be useful to search for
optical emission near the leading edges of
the X-ray emitting wakes.

\bigskip
\bigskip
\noindent
We thank Joachim Tr\"umper for a useful correspondence,
and valuable information regarding the ROSAT observations.
A.~L.\ and J.~S.\ acknowledge support from NSF and NASA,
and thank the Institute for Advanced Study in Princeton for hospitality.
F.~A.~R.\ is supported by a Hubble Fellowship.

\vfill\eject

\centerline{\bf REFERENCES}
\noindent
1. Tr\"umper, J. in {\it Proc. of the 1992 Texas/PasCos Symposium:
Relativistic Astrophysics and Particle Cosmology\/}
(eds. Akerlof, C.\ W.\ \& Srednicki, M.\ A.) 260--270 (Annals of
the New York Academy of Sciences, New York, 1993).

\smallskip
\noindent
2. \"Ogelman, H., Koch-Miramond , L., \& Auri\'ere, M. {\it Astrophys.\ J.\
Lett.\/} {\bf 342},
  L83--L86 (1989).

\smallskip
\noindent
3. Bailes, M.\ {\it et al.\/} {\it Astrophys.\ J.\ Lett.\/} {\bf 343},
  L53--L55 (1989).

\smallskip
\noindent
4. Burrows, A. \& Fryxell, B.\ A. {\it Science\/} {\bf 258}, 430--434 (1992).

\smallskip
\noindent
5. M\"onchmeyer, R., Sch\"afer, G., M\"uller, E. \& Kates, R.\ E.
{\it Astron.\ Astrophys.\/} {\bf 246}, 417--440 (1991).

\smallskip
\noindent
6. Livio, M.\ \& Soker, N. {\it MNRAS\/} {\bf 208}, 763--781 (1984).

\smallskip
\noindent
7. Harrison, P.\ A., Lyne, A.\ G.\ \& Anderson, B. {\it MNRAS\/}
{\bf 261}, 113--124 (1993).

\smallskip
\noindent
8. Nakamura, T.\ \& Fukugita, M. {\it Astrophys.\ J.\/}
{\bf 337}, 466--469 (1989).

\smallskip
\noindent
9. Nakamura, T.\ \& Oohara, K. {\it Prog.\ Theoret.\ Phys.\/}
{\bf 86}, 73--88 (1991).

\smallskip
\noindent
10. Rasio, F.\ A., \& Shapiro, S.\ L. {\it Astrophys.\ J.\/} {\bf 401},
226--245 (1992).

\smallskip
\noindent
11. Colpi, M.\ \& Rasio, F.\ A. to appear in {\it Evolutionary Links in the Zoo
of Interacting Binaries\/} (Mem.\ Soc.\ Astron.\ Ital., 1994).

\smallskip
\noindent
12. Woosley, S.\ E.\ \& Weaver, T.\ A. {\it Ann.\ Rev.\ Astron.\ Astrophys.\/}
{\bf 24}, 205--253 (1986).

\smallskip
\noindent
13. Lattimer, J.\ M., Mackie, F., Ravenhall, D.\ G.\ \& Schramm, D.\ N. {\it
Astrophys.\ J.\/} {\bf 213}, 225--233 (1977).

\smallskip
\noindent
14. Clayton, D.\ D. {\it Principles of Stellar Evolution
and Nucleosynthesis}, 275 (The University of Chicago Press, Chicago, 1983).

\smallskip
\noindent
15. Banit, M., Ruderman, M.\ A., Shaham, J.\ \& Applegate, J.\ H.
{\it
Astrophys.\ J.\/} {\bf 415}, 779--796 (1993).
\end